# Binding energies and radii of $N \geq Z$ nuclei in the n,p-networks model

I. Casinos

*Facultad de Química, Universidad del País Vasco, Pº Manuel de Lardizabal 3, 20018 San Sebastián, Spain*

e-mail: ismael.casinos@ehu.es

## Abstract

A nuclear model is extended to estimate binding energies and radii of neutron-rich nuclei. These calculations have been made for some representative examples of even-$Z$ and odd-$Z$ nuclei with nucleon numbers lower than sixty. A comparison of results with experimental data supports the value and development of the model in nuclear studies.

## 1 Introduction

Studies on the interaction between nucleons under the quantum mechanics formalism allow reaching knowledge on nuclear structure and behaviour of nuclear matter. However, it is difficult to be applied in solving the ab initio treatment of a many-body system for non-simple nuclei and because of that it becomes convenient the usage of approximations and development of semiclassical models [1-6].

Calculation of the nuclear binding energy is useful as a test to assess the goodness of a model approach to study the atomic nucleus [7-12]. In a previous report [13], I have suggested a nuclear potential and a model, in a non-quantum mechanics approach, to study the binding energy and radius of $N=Z-1$, $Z$, $Z+1$ nuclei with nucleon numbers lower than sixty. And this was taken as a base to study the fusion reaction of two nuclei.

The study of neutron-rich nuclei has attracted the attention of researchers [14-23] in order to reach an understanding of the peripheral structure of these nuclei.

The purpose of the present work is to make use of the cited model to estimate binding energies and radii of several $N \geq Z$ nuclei.

## 2 Brief outline of the model

In this section, a partial account of the model is accomplished. A complete picture can be found in Ref. [13].

In order to estimate the binding energy of nucleus C, $E_b(C)$, the whole formation process from its nucleons $Zp + Nn \rightarrow C(Z, N)$ is considered by means of the successive and appropriate nucleon-additions $A+n \rightarrow C$ and $A+p \rightarrow C$ following Scheme 1 with the associated addition (separation) energies $S_n$ and $S_p$, respectively. In other words, a nucleus C formed in a particular nucleon addition

$$^1_0n \searrow \quad ^3_1H \searrow \quad ^5_2He \searrow \qquad \qquad \qquad ^{59}_{29}Cu$$
$$\quad \nearrow\; ^2_1H \;\dashrightarrow\; ^4_2He \;\dashrightarrow\; \searrow \quad ^6_3Li \;\dashrightarrow\; \dashrightarrow\; ^{58}_{29}Cu \;\dashrightarrow\;$$
$$^1_1H \nearrow \quad ^3_2He \nearrow \quad \searrow\; ^5_3Li \qquad \qquad \qquad \searrow\; ^{59}_{30}Zn$$

$4x+1 \quad 4x+2 \quad 4x+3 \quad 4x+4 \quad 4x+1 \quad 4x+2 \qquad 4x+2 \quad 4x+3$

$\dashrightarrow$      neutron-addition process

$\rightarrow$      proton-addition process

$x = 0, 1, 2, \ldots 14$.

to a nucleus A is held as nucleus A taking part in the next one.

$$S_n = -\frac{kM_A(R_A + R_n)^6}{8R_A^3 R_n^3 D_C^6} \tag{1}$$

$$S_p = -\frac{kM_A(R_A + R_p)^6}{8R_A^3 R_p^3 D_C^6} + \frac{K_e e^2 Z_A}{D_C} \tag{2}$$

$M_A$ and $Z_A$ are nucleon and proton numbers of nucleus A, respectively. $R_n = R_p = 1.1498$ fm are neutron and proton radii. The diffuse surface radius of nucleus A, $R_A = R_{0A} M_A^{1/3}$, is the distance where its nucleonic distribution starts to be affected by the interaction with an approaching nucleon (or nucleus). The kernel radius of nucleus C, $D_C = aM_A^{1/3} + b$ (where $M_A = M_C - 1$), denotes the location distance, related to an energy minimum, of the outermost nucleon constituting the globular nucleus in its ground state. Values for all these parameters are displayed in Table 1 to be appropriately applied in eqs. (1) and (2) for each successive nucleon addition along a nucleus C formation.

**Table 1.** Parameter values to be used in eqs. (1) and (2).

| Type of nuclide A | Type of nuclide C | $a$ (fm) | $b$ (fm) | $R_{0A}$ (fm) | $k$ (MeV fm$^6$) |
|---|---|---|---|---|---|
| $4x+4$ | $4x+1$ | 0.68994 | 1.5116 | 2.9271 | 4.8798 |
| $4x+1$ | $4x+2$ | 0.71841 | 1.3948 | 1.1858 | 41.129 |
| $4x+2$ | $4x+3$ | 0.87530 | 0.82888 | 2.2436 | 11.778 |
| $4x+3$ | $4x+4$ | 0.92065 | 0.64997 | 1.2772 | 43.113 |

## 3 Results of binding energy and radius

As indicated previously [13], the binding energy and radius of $N=Z-1$, $Z$, $Z+1$ nuclei with nucleon numbers lower than 60 were calculated in this model. At present, it is intended to do those calculations for nuclei in that range studying several exemplary nuclei showing $N \geq Z$.

This model regards a nucleus as formed by two interpenetrated similar networks of neutrons and protons displaying a dependence of the interaction strength, *k*, (see Table 1) on the *np*, *nn*, *pp* pairings between nucleons that favours the formation of compact α-particle clusters jointly to residual valence nucleons.

The assumed nucleonic distribution in $N \geq Z$ nuclei is described by taking two illustrative examples: the oxygen isotopes as even-*Z* nucleus and the fluorine isotopes as odd-*Z* nucleus.

The alternate addition of eight neutrons and eight protons following Scheme 1 forms four α-particle clusters composing the nuclide $^{16}_{8}O$. The next successive addition of six neutrons to constitute the corresponding isotopes from $^{17}_{8}O$ to $^{22}_{8}O$ is made with the lowest interaction strength ($k$ = 4.8798 MeV fm$^6$) and each neutron makes use of a consecutive α-particle space.

In the case of fluorine isotopes, four α-particle clusters together the incomplete 5$^{th}$ one, made of both 9$^{th}$ neutron and 9$^{th}$ proton, form the nuclide $^{18}_{9}F$. The addition of the 10$^{th}$ neutron keeps incomplete the 5$^{th}$ α-particle with three nucleons in the nuclide $^{19}_{9}F$. And the successive addition of other five neutrons is made with the lowest interaction strength to form the corresponding isotopes from $^{20}_{9}F$ to $^{24}_{9}F$ where each neutron makes use of one consecutive α-particle space, as previously suggested for oxygen isotopes.

Calculations of binding energy per nucleon are shown in Fig. 1 and those of kernel radius in Fig. 2. And these results are displayed in Fig. 1(a) and Fig. 2(a) for even-*Z* nuclei (isotopes of helium, oxygen and sulphur), and in Fig. 1(b) and Fig. 2(b) for odd-*Z* nuclei (isotopes of lithium, fluorine and chlorine).

It can be seen in Fig. 1 that the calculated binding energies for nuclei of higher *Z* values are better adapted to experimental data [24] than nuclei of lower *Z* values.

In a high-*Z* nucleus the ratio of nucleons in its high-density core to neutrons in its low-density skin is higher than the ratio in a low-*Z* nucleus. A variation in a uniform nuclear density could affect the spherical radius and the interaction strength between two nuclear species that have not been taken into account in this work.

In conclusion, it is appreciated a concurrence between calculated and experimental data that sustains the usefulness of this model in nuclear studies.

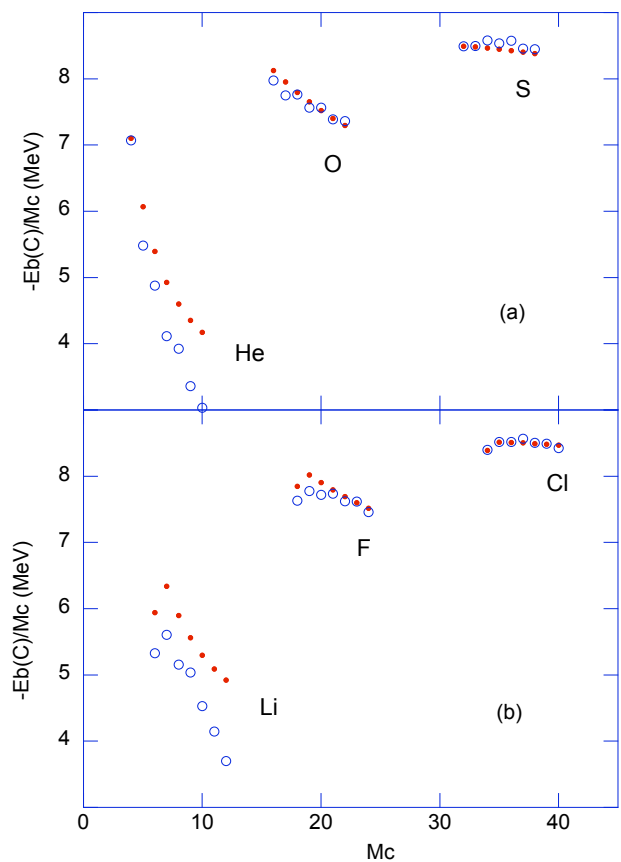

**Fig.1.** Calculated (filled circles) and experimental (unfilled circles) binding energies per nucleon for $N \geq Z$ nuclei. (a), even-$Z$ nuclei. (b), odd-$Z$ nuclei.

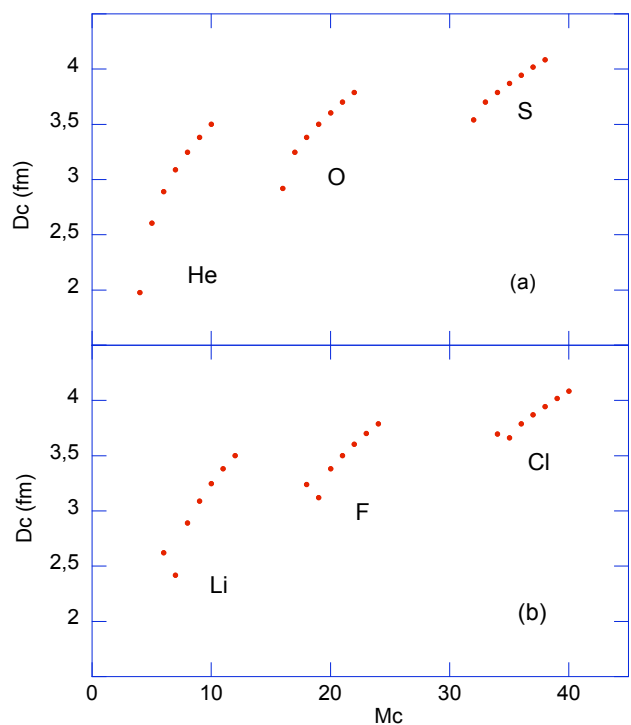

**Fig.2.** Calculated kernel radii for *N* ≥ *Z* nuclei. (a), even-*Z* nuclei. (b), odd-*Z* nuclei.